\def\year{2022}\relax
\documentclass[letterpaper]{article}\usepackage[]{graphicx}\usepackage[]{color}
\makeatletter
\def\maxwidth{ %
  \ifdim\Gin@nat@width>\linewidth
    \linewidth
  \else
    \Gin@nat@width
  \fi
}
\makeatother

\definecolor{fgcolor}{rgb}{0.345, 0.345, 0.345}

\usepackage{framed}
\makeatletter
 {\par\unskip\endMakeFramed%
 \at@end@of@kframe}
\makeatother

\definecolor{shadecolor}{rgb}{.97, .97, .97}
\definecolor{messagecolor}{rgb}{0, 0, 0}
\definecolor{warningcolor}{rgb}{1, 0, 1}
\definecolor{errorcolor}{rgb}{1, 0, 0}

\usepackage{alltt} 
\usepackage{aaai21_copyright}  
\usepackage{times}  
\usepackage{helvet} 
\usepackage{courier}  
\usepackage[hyphens]{url}  
\usepackage{graphicx} 
\usepackage{natbib}  
\usepackage{caption} 
\definecolor{c77a1d2}{RGB}{119,161,210}
\definecolor{bf9837}{RGB}{191,152,55}
\definecolor{cc0c0c0}{RGB}{192,192,192}

\usepackage{xcolor}
\definecolor{mycomp}{HTML}{fac631}
\definecolor{mymut}{HTML}{0d0887}

\usepackage{tikz}
\usepackage{booktabs}
\usepackage{multicol}
\AtBeginDocument{%
  \providecommand\BibTeX{{%
    \normalfont B\kern-0.5em{\scshape i\kern-0.25em b}\kern-0.8em\TeX}}}
    
\usepackage{subcaption}
\def\citepos#1{{\citeauthor{#1}}'s}
\usepackage[htt]{hyphenat}
\usepackage{amsthm}
\usepackage{amsmath}
\usepackage{commath}
\usepackage{mathtools}
\usepackage{amsmath}
\renewcommand{\widetilde}[1]{\mathbin{%
    \stackrel{\sim}{\smash{#1} \rule{0pt}{1.15ex}}%
    }}

\let\norm\undefined 
\DeclarePairedDelimiter\norm{\lVert}{\rVert}

\def\Slash{\slash\hspace{0pt}}

\title{Identifying Competition and Mutualism Between Online Groups}

\author{
 Nathan TeBlunthuis \textsuperscript{\rm  1 2},
 Benjamin Mako Hill \textsuperscript{\rm 1}\\}
 
 \affiliations{\textsuperscript{\rm 1} Department of Communication, University of Washington, \\
 \textsuperscript{\rm 2} Department of Communication Studies, Nothwestern University \\
 nathante@uw.edu}

\IfFileExists{upquote.sty}{\usepackage{upquote}}{}
\begin{document} 

\maketitle


\begin{abstract}
Platforms often host multiple online groups with overlapping topics and members. How can researchers and designers understand how related groups affect each other? Inspired by population ecology, prior research in social computing and human-computer interaction has studied related groups by correlating group size with degrees of overlap in content and membership, but has produced puzzling results: overlap is associated with competition in some contexts but with mutualism in others. We suggest that this inconsistency results from aggregating intergroup relationships into an overall environmental effect that obscures the diversity of  competition and mutualism among related groups. 
Drawing on the framework of community ecology, we introduce a time-series method for inferring competition and mutualism. 
We then use this framework to inform a large-scale analysis of clusters of subreddits that all have high user overlap. We find that mutualism is more common than competition.
\end{abstract}

\maketitle




\section{Introduction}
\label{sec:intro}

Online groups do not exist in isolation.\footnote{We use the term ``online group'' instead of ``online community'' to help avoid confusion with our term ``community ecology'' which plays an important conceptual and analytic role in our paper.}  Analyzing interdependence between online groups can be complex \citep{hill_studying_2019}, but recent social computing research has sought to quantify how online groups share users or topics \cite{datta_identifying_2017}, and how such interactions relate to outcomes like the emergence of new groups \cite{tan_tracing_2018}, the spread of hate speech \cite{chandrasekharan_you_2017} and contributions to peer-produced knowledge \cite{vincent_examining_2018}.  While this work has demonstrated that intergroup interactions matter, very little intergroup research has tackled questions of group success---i.e., why some online groups succeed in maintaining active and long-lived participation while most do not.
Can intergroup relationships 
explain whether online groups will grow or decline?





We seek to answer this question in terms of \emph{ecological relationships} following prior studies in social computing \cite{wang_impact_2012, zhu_impact_2014, zhu_selecting_2014}, we take inspiration from organizational ecology, an influential body of theory in sociology, and analyze \emph{competition} and \emph{mutualism} between online groups
\cite{hannan_organizational_1989, baum_ecological_2006}. 
Prior ecological studies of online groups have yielded inconsistent results that differ both from one context to another and from theoretical predictions. 
For example, wikis whose memberships overlap with other wikis survived longer \cite{zhu_selecting_2014}, but Usenet groups with overlapping memberships failed more quickly \cite{wang_impact_2012}. 

We propose that limitations of the \emph{population ecology} framework used by these studies give rise to these inconsistencies. We therefore introduce an alternative framework inspired by \emph{community ecology} that seeks to directly study competitive and mutualistic interactions between groups.
Population ecology models how overlapping resources among groups affect their subsequent growth, decline, or survival \citep{freeman_community_2006}, but it does not directly study interactions. 
By contrast, community ecology models related groups as an ``ecological community''  structured by a network of competitive and mutualistic relationships.

We introduce our community ecology approach and compare it to the population ecology approach from prior work in a two-part empirical study of 641 clusters of online groups among the 10,000 communities on Reddit with the most contributors. 
In Study A, we demonstrate a prototypical population ecology analysis by testing density dependence theory.
Prior studies would interpret the results of this analysis as suggesting that high degrees of user overlap are associated with competition. 

In Study B, we introduce our method for inferring networks of ecological relationships among related online groups based on clustering analysis and vector autoregression (VAR) models of group size over time \cite{ives_estimating_2003}. We illustrate the method in four case studies and present a large-scale analysis showing that mutualistic interactions are far more common than competitive ones.  To validate our approach, we show that including ecological interactions in our VAR models improves time series forecasting. 




Our findings illuminate the different contributions of population ecology and community ecology.
While Study A suggests that competition is strongest when user overlap is high, Study B finds widespread mutualism among groups with highly overlapping memberships.
Although these findings might seem contradictory, population and community ecology analyses provide complementary views. Population ecology points to favorable or unfavorable conditions for building online groups---conditions that may or may not involve competition and mutualism.  A community ecology analysis can infer local networks of competition and mutualism to explain how specific ecological relationships contribute to growth or decline. By demonstrating that ecological relationships within clusters of highly related groups are important---and by describing how to measure them---this paper lays the groundwork for future investigations into interdependent online groups and designs that support ecological communities.

\section{Related Work}
\label{sec:related.work}

Online groups are important sites for social support \cite{de_choudhury_mental_2014}, entertainment \cite{ducheneaut_alone_2006}, information sharing \cite{benkler_wealth_2006}, and political mobilization of disinformation campaigns and protest movements \cite{choudhury_social_2016, benkler_social_2013, krafft_disinformation_2020}.
Although an online group's ability to achieve its goals depends on attracting and retaining contributors, few develop a sizable group of participants \cite{kraut_building_2012}. Many attempts to explain growth and decline of online groups look to properties of individual groups like characteristics of founders and designs for regulating behavior \cite{kraut_building_2012, halfaker_rise_2013}.


By contrast, recent research shows the importance of interdependence among online groups \cite{kairam_life_2012, tan_tracing_2018, waller_generalists_2019}. 
For example, banning hate subreddits reduced hate speech in related subreddits \cite{chandrasekharan_you_2017}, Reddit and Stack Overflow receive substantial benefits from activity on Wikipedia \cite{vincent_examining_2018}, and editors make valuable and qualitatively different contributions across different languages of Wikipedia \cite{hale_cross-language_2015}. 
Our work contributes to this literature by providing a new conceptual lens and statistical method for studying intergroup connections. 


\subsection{Ecological Interdependence}
\label{sec:rdp}

Ecological approaches to online groups see online groups as depending on resources. Our conceptual approach, like prior ecological research in social computing and information systems, builds on resource dependence theory (RDT) \cite{butler_membership_2001, wang_impact_2012}. 
According to RDT, members of online groups contribute resources such as content, information, attention or social interactions that sustain the group.  
Ecological approaches observe that interrelated online groups may share resources with one another and effect each other's growth and survival as a result.  \textit{Rival resources} like participants' time, attention, and efforts become unavailable to others when used by one group \cite{benkler_wealth_2006, romer_endogenous_1990}, and competition over important rival resources can explain declines participation \cite{wang_impact_2012}. 



On the other hand, the value of a \textit{nonrival} resource does not decrease (and may even increase) when it is used. Nonrival resources that ``spill over'' can result in mutualism that promotes growth in related groups   \cite{zhu_impact_2014}.   
For example, the usefulness of a communication network increases as more people join it \cite{fulk_connective_1996}. Similarly, the usefulness of an information good can increase as more people come to know, refer to, and depend upon it.
Ecological approaches seek to understand how different types of resources will limit or promote growth.

\subsection{Population Ecology, Density Dependence and Overlapping Resources}
\label{sec:ecology_background}


Our work builds on a tradition rooted in \textit{organizational ecology}. First developed in the late 1970s by sociologists studying interactions between firms, organizational ecology was inspired by, and has drawn closely from, ecological studies in biology \citep{hannan_organizational_1989}. 
Organizational ecology has inspired at least three high-quality empirical studies of how resources shared by online groups shared shape their growth, decline, or survival \cite{wang_impact_2012, zhu_impact_2014, zhu_selecting_2014}.
These studies draw from the \textit{population ecology} strand of organizational ecology, and specifically enage with \textit{density dependence theory} (DDT).




DDT concieves of competitive or mutualistic forces as a function of population \textit{density}. In the earliest and most influential studies of DDT, density is simply the size of the population, which defines a homogenous set of organizations of groups facing the same competitive and mutualistic pressures \cite{aldrich_organizations_2006}.
However, online groups sharing a platform have diverse topics \cite{kairam_life_2012}, norms \cite{chandrasekharan_internets_2018}, and user bases \cite{tan_tracing_2018}. 
To account for this diversity, ecological studies of online groups have modeled density dependence based on the concept of \emph{overlap density} \cite{baum_ecological_2006,  wang_impact_2012, zhu_impact_2014, zhu_selecting_2014}. 
Overlap density measures the extent to which one group's members or topics overlap with all other groups'. Overlap density thus characterizes a group's \emph{niche} or local \emph{resource environment} defined by its distinctive topic and membership.



DDT proposes a model for the growth of organizational populations in which mutualism drives a virtuous cycle of population growth \cite{carroll_density_1989,hannan_organizational_1989}.
For example, a population of online groups, such as those sharing a platform, may grow in size as their platform gains in popularity, as established groups spin off new ones, and as useful knowledge develops that can be shared between groups \cite{tan_tracing_2018, zhu_impact_2014}.
On the other hand, when density is high, competition among population members over rival resources limits growth \cite{hannan_organizational_1989}. 
DDT thus proposes a trade-off in which low density reflects limited opportunities for mutualistic contributions of nonrival resources,  while high density reflects competition over rival resources.  
Therefore, DDT predicts that the relationship between density and positive outcomes like growth or survival is  $\cap$-shaped (inverse-U-shaped) \citep{baum_ecological_2006, carroll_density_1989}.

Tests of DDT in populations of online groups yield inconsistent results. In \citet{wang_impact_2012}, user overlap in Usenet newsgroups is associated with decreasing numbers of participants. Similarly, \citet{teblunthuis_population_2020} find that topical overlaps between online petitions are negatively associated with participation. 
By contrast, \citet{zhu_impact_2014} find that membership overlap is positively associated with increasing survival of new Wikia wikis. Only \citet{zhu_selecting_2014} find support for the $\cap$-shaped relationship predicted by DDT in an enterprise social media platform.

In Study A, we provide a test of DDT using data from Reddit. The classical logic of DDT appears reasonable in the context of Reddit because low overlap density likely reflects an impoverished environment lacking in non-rival resources like skills and knowledge of experienced users, while a group with high overlap density likely faces competition over its members \cite{zhu_selecting_2014, zhu_impact_2014}:
\textit{(\textbf{H1}) The relationship between overlap density and the growth of online groups is  $\cap$-shaped (inverse-U-shaped).}


\subsection{Introducing Community Ecology \label{sec:community_ecology}}

The distinction between population ecology and community ecology theories is in where they locate ecological dynamics like competition and mutualism.  In population ecology, competition and mutualism are properties of an environmental niche; but in community ecology, they are relations in networks of interdependent groups called \emph{ecological communities} \cite{freeman_community_2006}.
While most community ecology studies of classical organizations analyze ecological communities of different organizational forms, some, like our study, analyze communities of related organizations \cite{freeman_community_2006}. 



Community ecology focuses on \emph{ecological interactions} \cite{aldrich_organizations_2006}.
Mutualism is an ecological interaction where one group has a positive influence on the second such that growth in the first group leads to growth in the second.  Competition is when one group has a negative effect on the second such that growth in the first group leads to decline in the second. 
These relationships are modeled as edges of a directed network. As a result,
ecological interactions can be mutualistic in one direction and competitive in the other and mutualism (or competition) from one group to another group may (or may not) be returned in kind.  The goal of many community ecology analyses in 
both biology and organization science is to infer and analyze the \emph{community matrix}, which quantifies competitive and mutualistic interactions \cite{ives_estimating_2003, aldrich_organizations_2006}. 

In Study B, we demonstrate community ecology by inferring networks of ecological interactions in ecological communities on Reddit to determine whether mutualism or competition among subreddits is more common. We then present case studies to illustrate different types of ecological communities.
Finally, we evaluate whether modeling ecological interactions is useful for making time series forecasts of participation in online groups:
(\textit{\textbf{H2}) The addition of ecological interactions to a baseline time series model improves the forecasting performance.}

\section{Materials \& Methods}
\label{sec:methods}



We analyze data from the publicly available Pushshift archive of Reddit submissions and comments from December 5\textsuperscript{th} 2005 to April 13\textsuperscript{th} 2020
\cite{baumgartner_pushshift_2020}. Within this dataset, we limit our analysis to submissions and comments from the 10,000 subreddits with the highest number of comments. There are 702 subreddits larger than the smallest subreddit included in our dataset having a majority of submissions marked ``NSFW,'' which typically indicates pornographic material. As others have done in large-scale studies of Reddit \cite[e.g.,][]{datta_identifying_2017}, we exclude these subreddits to avoid asking members of our research team to inspect clusters including pornography. The top 10,000 subreddits provide a sufficiently large number of ecological communities for our statistical analysis. 

\subsection{Study A: Density Dependence Theory} 
\label{methods:density}

\noindent \textbf{User overlap} $o_{i,j}$ quantifies the degree to which two subreddits ($i$ and $j$) share users. 
\citet{zhu_impact_2014} and \citet{wang_impact_2012} both measure user overlap between two groups by counting the number of users contributing to both groups at least once and exclude users who appear in more than 10 groups. In our preliminary analysis, we found that this measure led to similarity measures and clusters with poor face validity.  These issues may have stemmed from how Reddit users often peripherally participate in many groups while participating heavily in few \citep{zhang_community_2017}. 
Therefore, our measure of user overlap follows \citet{datta_identifying_2017} by using the number of comments each user makes in each pair of groups.




To measure user overlap between subreddits, we first build user frequency vectors by counting the number of times each user comments in each subreddit. We prevent giving undue weight to subreddits with higher overall activity levels by normalizing the comment counts for each subreddit by the maximum number of comments by a single author in the subreddit:
\begin{equation}
    f_{u,j} = \frac{n_{\mathrm{u,j}}}{max_{v\in\mathrm{J}}n_{v,j}} \label{eq:user.frequency}
\end{equation}
 

\noindent where $n_{u,j}$,  the user frequency, is the number of times that user $u$ authors a comment in subreddit $j$. This results in a user frequency vector $F_j$ for each subreddit that is sparse and high-dimensional, having one element for each user account that comments in any subreddit in our dataset.
Next, we use latent semantic analysis (LSA) to reduce the dimensionality of the user frequency vectors: 
\begin{align}
    \mathbf{F} &= \mathbf{U \Sigma V}^T \\ \nonumber
    \widetilde{F_{j}} &= \mathbf{U_k}^TF_j \label{eq:user.frequency.svd}
\end{align}

\noindent where $\mathbf{F}$ is the matrix where columns are author frequency vectors $F_j$ and $\mathbf{U \Sigma V}^T$ is its singular value decomposition. Truncating the singular value decomposition to use only the first $k$ left-singular vectors gives $\mathbf{U_k}$. Left-multiplying a subreddit's author frequency vector by $\mathbf{U_k}$ transforms the high-dimensional author frequencies into $\widetilde{F_j}$, their approximation in the $k$-dimensional space. 
Our measure of \textit{user overlap} ($o_{i,j}$) is the cosine similarity between these vectors: 



\begin{equation}
    o_{i,j} = \frac{\widetilde{F_{j}} \cdot \widetilde{F_{i}}} {\norm{\widetilde{F_i}} \norm{\widetilde{F_j}}} \label{eq:user.overlap}
\end{equation}

\noindent where $\norm{\widetilde{F_i}}$ is the euclidean norm of the transformed user frequencies for subreddit $i$.




\noindent \textbf{Growth} $Y_i$, the dependent variable in our density dependence model testing H1 is measured as the change in the (log-transformed) size of a subreddit over the final 24 weeks of our data, from to November 4\textsuperscript{th} 2019 to April 13\textsuperscript{th} 2020. 
\vspace{0.8em}

\noindent \textbf{Overlap density} $d_i$, the normalized average user overlap for a given subreddit is the independent variable in our density dependence model testing H1:
\begin{align}\label{eq:user.overlap.density}
  d^*_{i} &= \frac{1}{\left|S\right|-1} \sum_{j\in R;j\ne i} \mathrm{o}_{i,j} \nonumber \\
  d_{i} &= \frac{d_i^*}{\mathrm{max}_j d_j^*}
\end{align}
\noindent where $S$ is the set of groups in our dataset.

\subsubsection{Regression model for H1} \label{sec:reg.H1}
To test H1, we fit Model 1:
\begin{align}
\mathrm{Model~1} & & Y_i = B_0 + B_1 d_{i} + B_2 d^2_{i}  \label{eq:M1}
\end{align}

\noindent where $Y_i$ is the growth of subreddit $i$ and $d_i$ is its overlap density. The model has first and second-order terms for overlap density to allow for a curvilinear relationship between \emph{overlap density} and \emph{growth}.

\subsection{Study B: Introducing Community Ecology}


\label{sec:clustering}
Analyzing networks of ecological interactions is the key difference between community ecology and population ecology. 
To identify ecological communities of related subreddits, we use a clustering procedure based on user overlap.
We selected a clustering model using grid search to obtain a high silhouette coefficient. The silhouette coefficient captures the degree to which a clustering creates groups of subreddits with high within-cluster similarity.

We ran the affinity propagation, HDBSCAN and \textit{k}-means clustering algorithms and selected the algorithm, hyperparameters, and LSA dimensions $k$ that resulted in the clustering with a high silhouette coefficient having less than 5,000 isolated subreddits, and at least 50 clusters. 
We limit the number of isolated subreddits because some choices of hyperparamters for the HDBSCAN algorithm could improve the silhouette coefficient, but at the cost of greatly increasing numbers of isolated subreddits.  
Choosing a high limit to the number of isolates helps ensure that our clusters contain highly related communities. 
We chose an HDBSCAN clustering with 731 clusters, 4964 isolated subreddits, $k=600$ LSI dimensions, and a silhouette score of 0.48.    We exclude the isolated subreddits from our analysis. More details about our clustering selection process are found in the online supplement.

We evaluate the external validity of the chosen clustering using the purity evaluation criterion.
To do so, an undergraduate research assistant examined a random sample of 100 clusters including 744 subreddits.  By visiting the subreddits and using their own judgment, the assistant flagged subreddits that did not seem like a good fit for their assigned cluster. Using these labels and excluding 25 subreddits that have been deleted, made private, or banned, we calculated the purity of our clustering as 0.92. Therefore, we estimate that  92\% of subreddits belong to their assigned cluster.



\vspace{0.8em}
\noindent \textbf{Group size} is the dependent variable of the  models we use to infer ecological interactions. Measured as the number of distinct users commenting in a subreddit each week, group size quantifies the number of people who participate in a subreddit over time. Typical of social media participation data, group size is highly skewed---i.e., we transform it by adding 1 and taking the natural logarithm. 

\subsubsection{Inferring ecological interactions}
\label{sec:var}

The community matrix $\mathbf{\Phi}$ of ecological interactions can be inferred from time series data using vector autoregression models (VAR models). VAR models are a workhorse in biological ecology because VAR(1) models (i.e., VAR models with a single autoregressive term) have a close relationship with the Gompertz of population growth, a common theoretical model in ecology
\cite{ives_estimating_2003}. 

VAR(1) models generalize auto-regressive AR(1) models in time series analysis. Where AR(1) models predict the state of a single time series as a function of its previous values, VAR(1) models predict several time series as a function of each other's previous values \cite{ives_estimating_2003}:

\begin{equation}\label{eq:var1}
Y_t = B_0 + B_1t + \sum_{k \in K}A_k x_{k,t} + \sum_{j \in M}\Phi_{j} y_{j,t-1} + \epsilon_t
\end{equation}

\noindent where $Y_t$ is a vector containing the sizes of a set of online groups ($M$) at time $t$. $B_0$ is the vector of intercept terms and $B_1$ is the vector of linear time trends ($b_{1,j}$) for each community ($j$). $\Phi_{j}$ represents the influence of $y_{j,t-1}$, the size of the $j^{\mathrm{th}}$ online group at time $t-1$ on $Y_t$. $\Phi_{j}$ is a column of $\mathbf{\Phi}$, a matrix of coefficients in which the diagonal elements correspond to intrinsic growth rates (marginal to the trend) for each online group and the off-diagonal elements are intergroup influences, and $\epsilon_t$ is the vector of error terms

Additional time-dependent predictors ($x_{k,t}$) can be included in the vectors $X_{k}$ with coefficients $a_k$. Because subreddits are created at different times, growth trends must begin only after the subreddit is created. We use $X_{k}$ to introduce a  counter-trend during the period prior to the creation of subreddits so that each group's growth trend begins in the period the group is created. For each group $j$ created at time $t^0_j$ we fill $X_{j}$ with the sequence $[1,2,3,\ldots\ ,t^0_j-1,0,0,0,\ldots\ ]$. In other words, $X_{j}$ adds a counter-trend only during the period prior to the first comment in subreddit $j$. We fix the elements $a_{j,i}$ of $A_j$ equal to 0 unless $i=j$, so the counter trend only influences subreddit $j$. This effectively sets $a_{j,j}$ approximately equal to $-b_{1,j}$. 

We fit VAR(1) models using ordinary least squares as implemented in the \texttt{vars} \texttt{R} package to predict the group size each week over the history of each subreddit prior to November 4\textsuperscript{th} 2019. We hold out 24 weeks of data for forecast evaluation and fit our models on the remainder. To ensure that sufficient data is available for fitting the models, we exclude 946 subreddits and 89 clusters having less than 156 weeks of activity. 



\subsubsection{Characterizing ecological communities}
\label{sec:characterizing.ecological.communities}

In Study B, we interpret the community matrix $\mathbf{\Phi}$ as a directed network of ecological interactions, a \emph{competition-mutualism network} \citep{ives_estimating_2003}. Although the elements of $\mathbf{\Phi}$ correspond to direct associations between group sizes, ecological interactions can also be indirect. Consider 3 one-directional interactions between three groups ($a$, $b$, $c$) such that growth in $a$ predicts decreased growth in $b$ ($\phi_{a,b} < 0$), growth in $b$ predicts decreased growth in $c$ ($\phi_{b,c} < 0$), but $a$ and $c$ do not directly interact ($\phi_{a,c} \approx 0$).


This does not necessarily mean that groups A and C are independent. Rather, an exogenous increase in A predicts a decrease in B and thereby an eventual increase in C.  Such indirect relationships are analyzed by using impulse response functions (IRFs) to interpret a VAR model. 
In large VAR models containing many groups, the great number of parameters can mean that few specific elements of $\mathbf{\Phi}$ will be statistically significant, even as many weak direct relationships can combine into statistically significant impulse response functions (IRFs) \cite{ives_estimating_2003}. 

An IRF predicts how much each group's size would change in response to a sudden increase in the size of each other group: 

\begin{equation}
    \mathbf{\Theta_t} = \mathbf{\Theta_{t-1}}\mathbf{\Phi}, t = 1,2,... \label{eq:irf} 
\end{equation}

\noindent where $\mathbf{\Theta_t}$ is the impulse response function at time $t$.   $\mathbf{\Theta_0}$ is an $M$-by-$M$ identity matrix so our impulses represent a log-unit increase of 1 to each group. $\mathbf{\Theta_t}$ is a matrix with elements $\theta^t_{i,j}$ corresponding to the response of group $j$ to the impulse of group $i$.  


We use IRFs of our VAR(1) models to make our visualizations of example competition-mutualism networks in §\ref{sec:case.studies}. 
We compute the IRFs with bootstrapped confidence intervals (CI) based on 1,000 samples using the \texttt{vars} \texttt{R} package. We draw an edge $i \rightarrow j$ in the competition-mutualism network if the 95\% CI of $\theta^t_{i,j}$ does not include zero at any time $10>=t>0$.  If $\theta^t_{i,j} >0 $, the edge indicates mutualism and if  $\theta^t_{i,j} < 0$  the edge indicates competition.


\vspace{0.8em}
\noindent \textbf{Average ecological interaction}
$\overline{m}$ measures the extent to which an overall ecological community is mutualistic or competitive by taking the mean point estimate of the off-diagonal coefficients of $\mathbf{\Phi}$:
\begin{equation}\label{eq:average.interaction}
\overline{m} = \frac{1}{\left|M\right| - 1} \sum_{i\in M} \sum_{j\in M;j\ne i} \phi_{i,j}
\end{equation}

\noindent If $\overline{m} > 0$ then mutualistic interactions within the ecological community are stronger than competitive ones, and if $\overline{m} < 0$ then competitive interactions are stronger then mutualistic ones.


\vspace{0.8em}
\noindent \textbf{Ecological interaction strength}
$\kappa$ quantifies the overall strength of ecological interactions in an ecological community as the mean absolute value of the point estimates of the off-diagonal coefficients of $\mathbf{\Phi}$:
\begin{equation}\label{eq:average.absolute.interaction}
\kappa = \frac{1}{\left|M\right| - 1} \sum_{i\in M} \sum_{j\in M;j\ne i} \left| \phi_{i,j} \right|
\end{equation}

\noindent where $\left| \phi_{i,j} \right|$ is the absolute value of the coefficient $\phi_{i,j}$.
The average ecological interaction can be close to 0 if ecological interaction strength is low or if the ecological interaction strength is high and results from a mixture of competitive and mutualistic interactions that cancel one another out when averaged.

\subsubsection{Forecasting growth}
\label{sec:mes.forecasting}
To test H2, we evaluate whether modeling ecological interactions improves time series forecasting of future participation in online groups by comparing the model in Equation \ref{eq:var1} to a baseline model with  off-diagonal elements of $\mathbf{\Phi}$ fixed to 0. This baseline model is equivalent to our VAR model, but excludes ecological interactions.

We compare our VAR model to the baseline in terms of two forecasting metrics with differing assumptions: root-mean-square-error (RMSE) and the continuous ranked probability score (CRPS).  RMSE is commonly used, non-parametric, and intuitive, but does not take differing scales of the predicted variable or forecast uncertainty into account.  Thus,  it may place excessive weight on larger subreddits having greater variation in size. 
The CRPS accounts for variance in the data and rewards forecasts for both accuracy and precision and is thus a ``proper scoring rule'''' for evaluating probabilistic forecasts \cite{gneiting_strictly_2007}. 
Our CRPS calculations assume that the predictive forecast distribution for each community is normal with standard deviations given by the 68.2\% forecast confidence interval. We calculate CRPS using the \texttt{scoringRules} \texttt{R} package.

\section{Results}
\label{sec:results}

\subsection{Study A: Density Dependence Theory}
\label{sec:res:studyA}

We test density dependence theory as formulated in H1 using Model 1 
which has first- and second-order terms for the effect of overlap density on growth.  As described in §\ref{sec:ecology_background}, H1 hypothesizes that overlap density will have a curvilinear $\cap$-shaped (inverse-U-shaped) relationship with growth indicated by a positive first-order regression coefficient and a negative second-order coefficient.

\begin{figure}
  \centering

\includegraphics[width=\columnwidth]{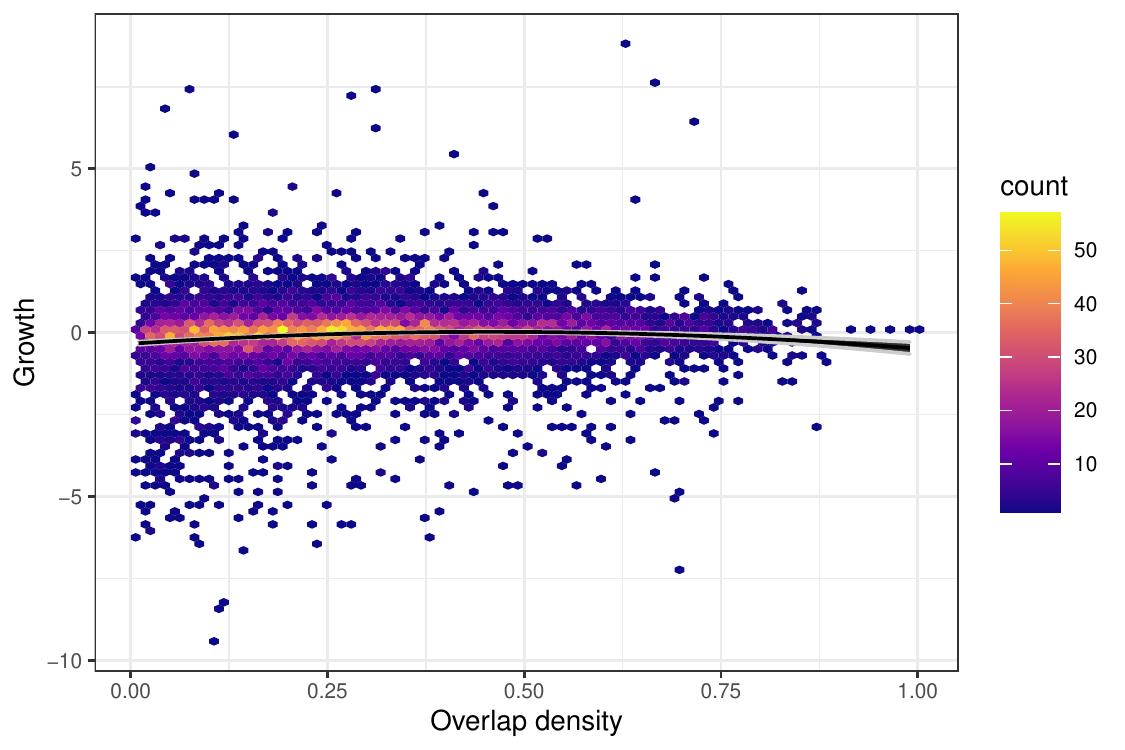} 

\caption{A 2D histogram of subreddits with overlap density (log-transformed) on the X-axis and the change in the logarithm of the number of distinct commenting users on the Y-axis.  The black line shows the marginal effect of overlap density on growth as predicted by Model 2. The gray region shows the 95\% confidence interval of the marginal effect. \label{fig:density}}
\end{figure}



We observe this predicted relationship between overlap density and growth. Figure \ref{fig:density} plots the marginal effects of overlap density on growth for the median subreddit laid over a scatterplot of the data.
The point where increasing density ceases to predict increasing growth and begins to predict decreasing growth is at the 49\textsuperscript{th} percentile. 
Prototypical subreddits at this overlap density grew slightly (95\% CI:[0.001,0.06]).  Yet subreddits at the lower and upper extremes of overlap density slightly declined on average. Typical groups at the 20\textsuperscript{th} percentile of overlap density decline by 1.1 members (95\% CI:[-1.1,-1.15]) and typical groups at the 80\textsuperscript{th} percentile decline by 1.2 members (95\% CI:[-1.1,-1.28]). 


\subsection{Study B: Introducing Community Ecology}
\label{sec:res.characterizing}

\begin{figure}

\includegraphics[width=\columnwidth]{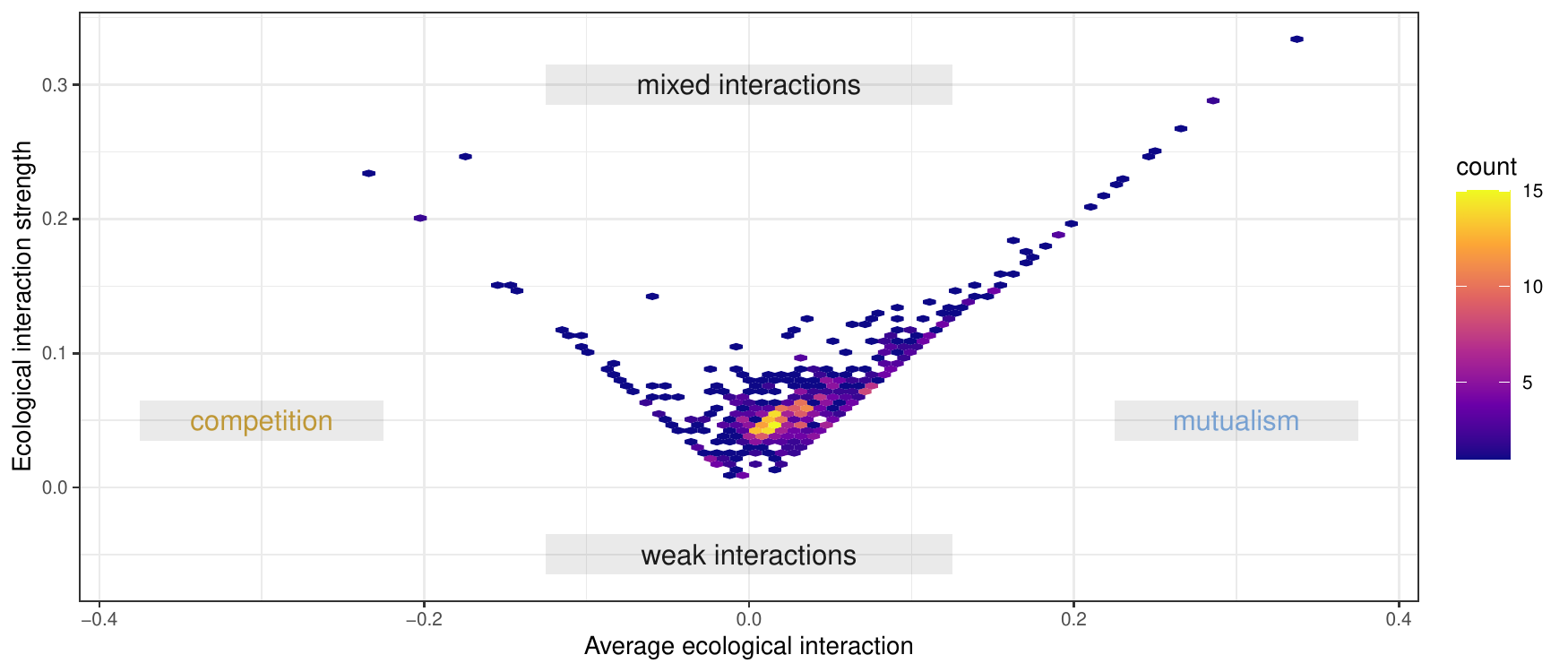} 

\caption{Two-dimensional histogram showing ecological communities on Reddit in our typology.  The X-axis shows the overall degree of mutualism or competition in clusters of subreddits with high user overlap based on the average ecological interaction.  The Y-axis shows the ecological interaction strength representing the overall magnitude of competition or mutualism.}

\label{fig:commense.x.abs.commense}
\end{figure}

Figure \ref{fig:commense.x.abs.commense} visualizes the distribution of average ecological interaction and ecological interaction strength over the 641 ecological communities we identify.  
We observe ecological communities characterized by strong forms of both mutualism and competition, others having mixtures of the two, and some with few significant ecological interactions.  Mutualism is more common than competition, with the mean community having an average ecological interaction of 0.03 ($t=14.5$, $p<0.001$). We find that 524 clusters (81.7\%) are mutualistic. Not only are most ecological communities mutualistic, but more mutualistic ecological communities have greater ecological interaction strength (Spearman's $\rho=0.58$, $p<0.001$).
Therefore, our community ecology analysis suggests that among groups with similar users, mutualistic ecological interactions are more common than competitive ones.

\subsubsection{Example ecological communities}
\label{sec:case.studies}

We present four case studies to illustrate our typology of ecological communities of online groups. Figure \ref{fig:commense.x.abs.commense} shows that we find clusters of subreddits characterized by mutualism, competition, a mixture of mutualism and competition, and few ecological relationships at all. We select one case from each of these four types using our measures of average ecological interaction ($\overline{m}$) and ecological interaction strength ($\kappa$). To allow for more interesting network structures, we draw our cases from the 367 large clusters having at least five subreddits. 

\begin{figure*}
\begin{minipage}[t][][t]{0.49\columnwidth}
  \centering
  \includegraphics[width=1\columnwidth,trim=0.9in 0.9in 0.9in 0.9in]{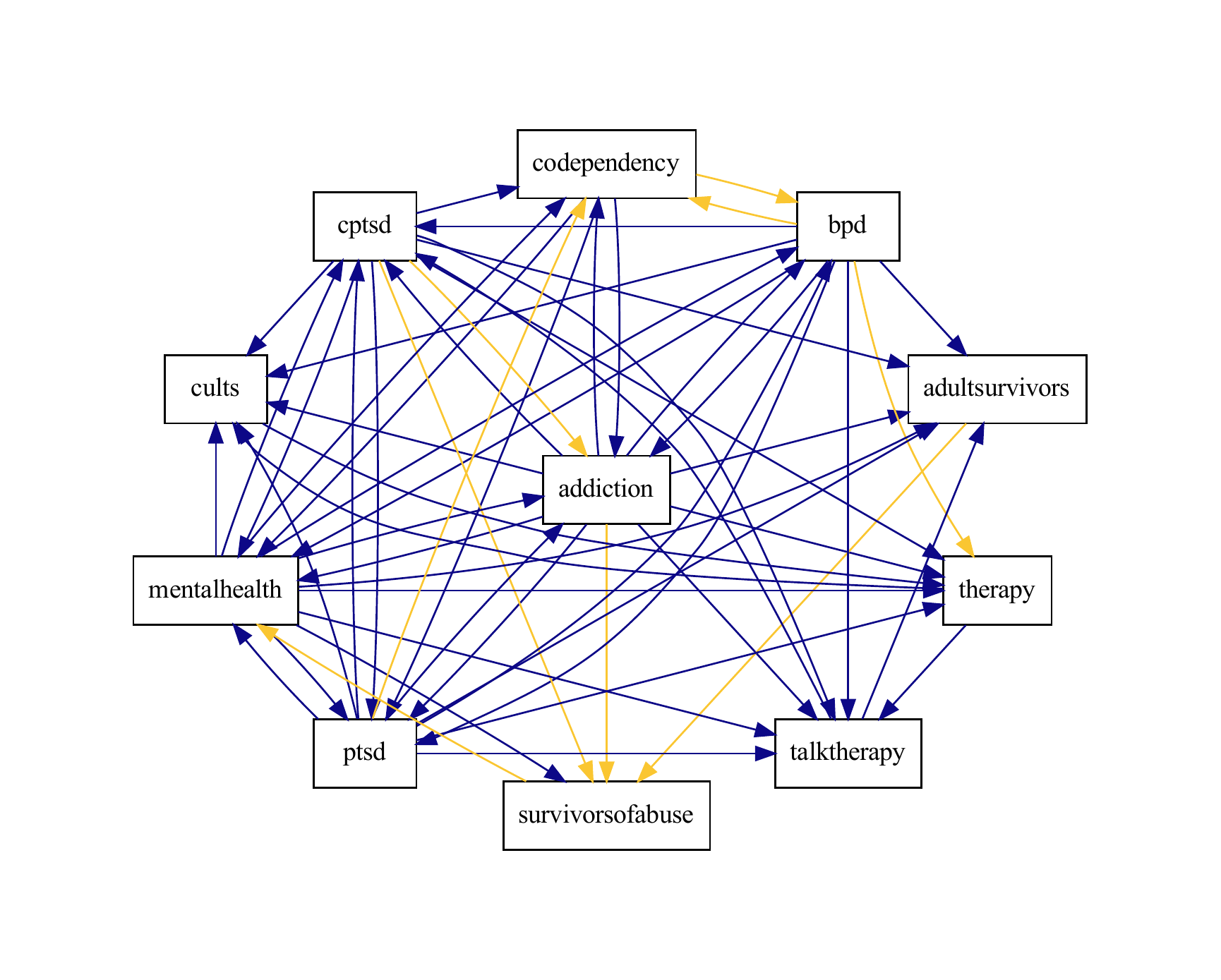}
  \subcaption{The ecological community of subreddits for supporting mental health and survivors of abuse is dense with largely mutualistic interactions. 
  \label{fig:mut.network}}
\end{minipage}
\hfill
\begin{minipage}[t][][t]{0.49\columnwidth}
  \centering
  \includegraphics[width=1\columnwidth,trim=0.9in 0.9in 0.9in 0.9in]{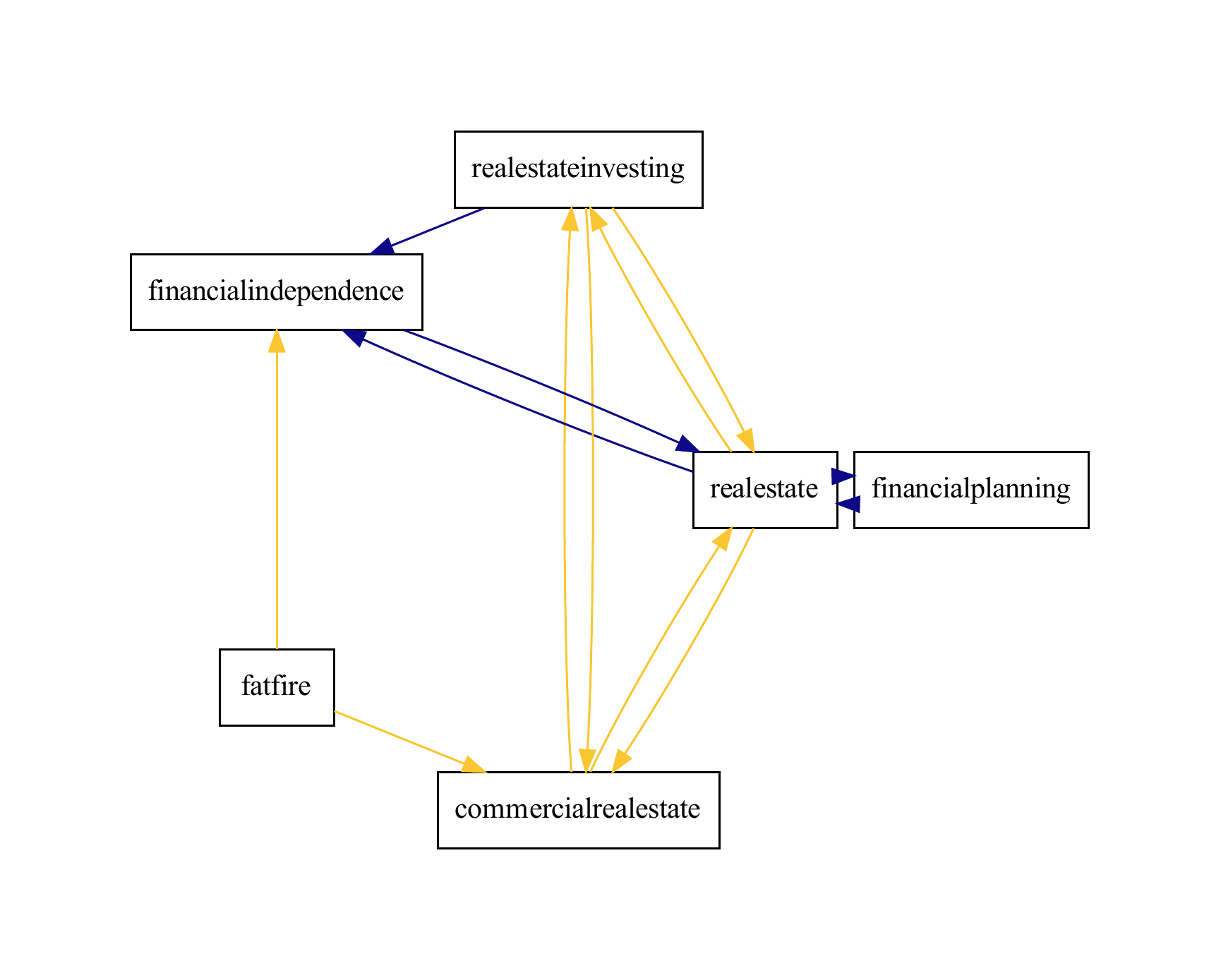}
  \subcaption{ The subreddits about real estate and finance are relatively competitive.
  \label{fig:comp.network}}
\end{minipage} \hfill
\begin{minipage}[t][][t]{0.49\columnwidth}
  \centering
  \includegraphics[width=1\columnwidth,trim=0.9in 0.9in 0.9in 0.9in]{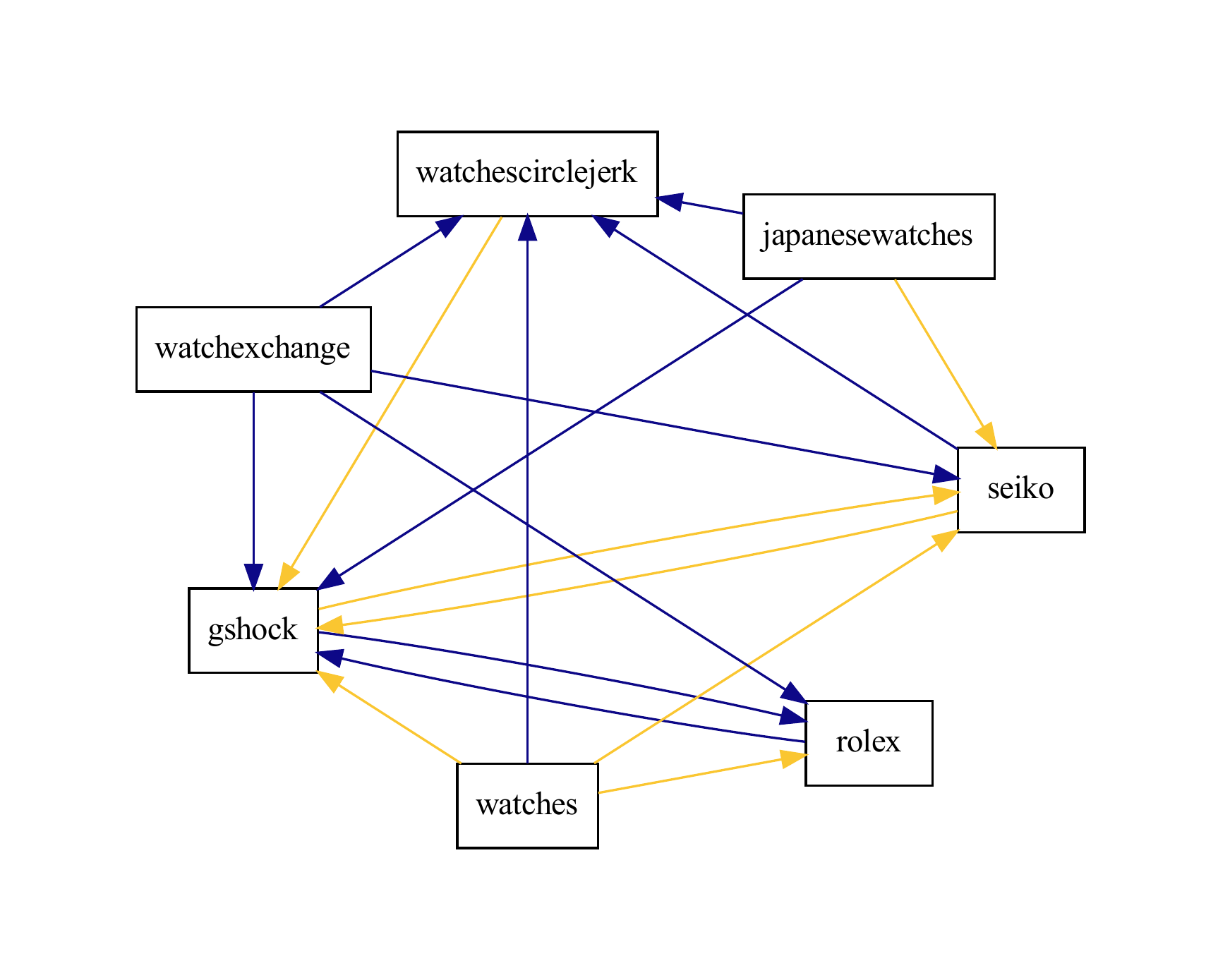}
  \subcaption{Subreddits about watches are dense with both mutualistic and competitive interactions. 
  \label{fig:mixed.network}}
\end{minipage}
\hfill
\begin{minipage}[t][][t]{0.49\columnwidth}
  \centering
  \includegraphics[width=1\columnwidth,trim=0.9in 0.9in 0.9in 0.9in]{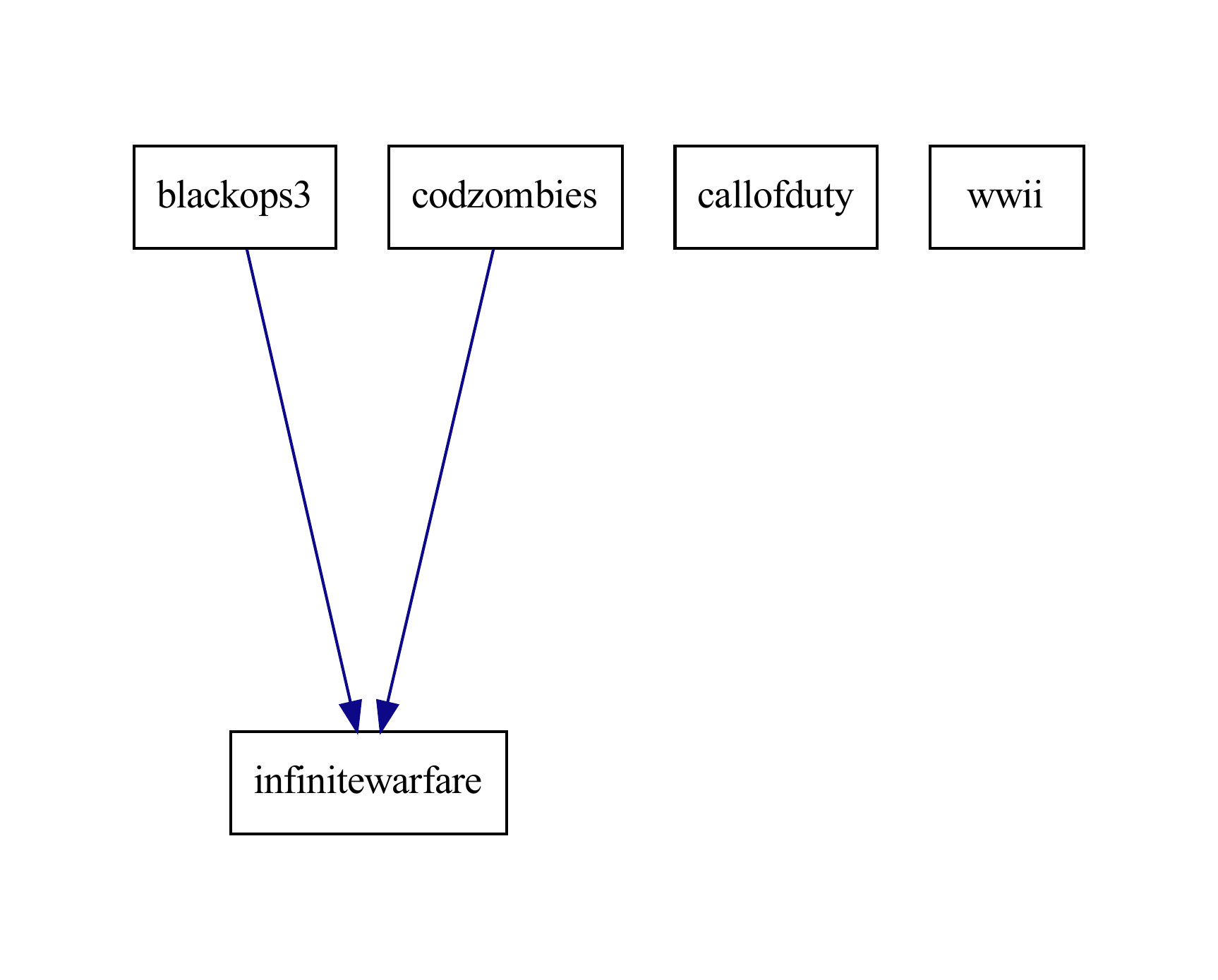}
  \subcaption{The ecological community of subreddits about Call of Duty video games is characterized by relatively sparse ecological interactions.
  \label{fig:void.network}}
\end{minipage}
\caption{Network visualizations of commensal relationships in example ecological communities of subreddits with overlapping users. \textcolor{mycomp}{Yellow} indicates \textcolor{mycomp}{competition} and \textcolor{mymut}{purple} indicates \textcolor{mymut}{mutualism}.  \label{fig:networks}}
\end{figure*}

Figure \ref{fig:networks} presents visualizations of competition-mutualism networks representing statistically significant impulse response functions. For each example, we examined the terms of the vector autoregression parameter $\mathbf{\Phi}$, the impulse response functions, and the model fits and forecasts, all of which are available in our online supplement.  We also visited each subreddit in the clusters and read their sidebars and top posts to support our brief qualitative descriptions.

\subsubsection{Mutualism among mental health subreddits}

To find a case characterized by mutualism, we selected the top 37 large clusters with the greatest average ecological interaction. From these, we arbitrarily chose one interesting ecological community, the \textit{mental health} cluster, which includes 11 subreddits for supporting people in struggles with mental health, addiction, and surviving abuse.  
Constitutive subreddits include those focused on specific mental health diagnoses like \texttt{r\Slash bpd} (bipolar disorder) and \texttt{r\Slash cptsd} (complex post traumatic stress disorder) while others like \texttt{r\Slash survivorsofabuse} and \texttt{r\Slash adultsurvivors}
are support groups. 

The interactions among these subreddits are dense and primarily mutualistic as shown in Figure \ref{fig:mut.network}. There are a handful of competitive interactions like the reciprocal competition detected between \texttt{r\Slash codedependence} and \texttt{r\Slash bpd}. We also observe some interactions that are mutualistic in one direction and competitive in the other. For example, growth in \texttt{r\Slash addiction} predicts increasing growth in \texttt{r\Slash cptsd} even as that growth in \texttt{r\Slash cptsd} predicts decreasing growth in \texttt{r\Slash addiction}. This suggests a pattern in which \texttt{r\Slash cptsd} siphons members from \texttt{r\Slash addiction}. That said, the density of mutualistic interactions shown in Figure \ref{fig:mut.network} suggests that different subreddits have complementary roles in this ecological community as people turn to different types of groups for help with interrelated problems.  While attempting to explain why different online groups form mutualistic or competitive interactions is left to future research, the example of mental health subreddits demonstrates how groups with related topics and overlapping participants can have mutualistic interactions.

\subsubsection{Competition among real estate and finance subreddits}

To find competitive clusters, we selected an ecological community that we label \textit{finance} from the 36 large clusters with the lowest average ecological interaction having six subreedits. Three of them: \texttt{r\Slash realestateinvesting}, \texttt{r\Slash realestate} and \texttt{r\Slash commercialrealestate}, deal in different aspects of the real estate industry while \texttt{r\Slash financialindependence} and \texttt{r\Slash fatfire} (the acronym ``fire'' means ``financial independence/retire early'') focus on building wealth and becoming financially independent and \texttt{r\Slash financialplanning} is a general financial advice subreddit.

In contrast to the mental health ecological community, the finance cluster has mostly competitive ties as visualized in Figure \ref{fig:comp.network}. We detect three reciprocal competitive interactions among the three subreddits that focus on real estate. The edges from \texttt{r\Slash fatfire} to \texttt{r\Slash commercialrealestate} and \texttt{r\Slash financialindependence} are competitive as well.   

Although this cluster is among the most competitive in our data, it contains mutualistic ties between the general finance subreddits (\texttt{r\Slash financialplanning} and \texttt{r\Slash financialindependence}) and \texttt{r\Slash realestate}. This reflects just how prevalent mutualism is among subreddits with high degrees of user overlap. 


\subsubsection{Mixed interactions among timepiece subreddits}

Next, we turn to the \textit{timepiece} ecological community of
7 subreddits about watches that has low average ecological interaction but high ecological interaction strength. 
We selected the \textit{timepiece} subreddits from 36 
large clusters with the average ecological interaction  closest to 0 and then from the 15 clusters with the greatest ecological interaction strength.

As shown in Figure \ref{fig:mixed.network}, the network of timepiece subreddits is dense with ecological interactions (although not as dense as the mental health subreddits). We observe both reciprocated mutualistic interactions, like that between \texttt{r\Slash rolex} and \texttt{r\Slash gshock}, and competitive interactions like that between \texttt{r\Slash gshock} and \texttt{r\Slash seiko}.  We also observe numerous unreciprocated competitive and mutualistic relationships like the mutualism between \texttt{r\Slash watchexchange} and \texttt{r\Slash watchcirclejerk}\footnote{The suffix is widely understood on Reddit to signify a jokey, meme, or satirical subreddit.}
and the competition between \texttt{r\Slash japanesewatches} and \texttt{r\Slash seiko}.
Though the average ecological interaction among these subreddits is near 0, our analysis reveals a complex ecological community with a mixture of competition and mutualism.   
 
\subsubsection{Sparse interactions among Call of Duty subreddits}

To find a case where ecological interactions are weak, we return to the group of the 36 
large clusters with the average ecological interaction closest to 0 but select from the 15 clusters within this group with the lowest ecological interaction strength. From these, we chose the \textit{Call of Duty} cluster containing five groups about the popular series of video games.

The Call of Duty ecological community is sparse, having only two significant ecological interactions among its 5 member groups. This ecological community includes subreddits about different editions of the series such as \texttt{r\Slash blackops3}, \texttt{r\Slash infinitewarfar} and \texttt{r\Slash wwii} as well as one about a popular spin-off zombie game \texttt{r\Slash codzombies} and the more general \texttt{r\Slash callofduty} subreddit. We find that  growth in \texttt{r\Slash blackops3} or \texttt{r\Slash codzombies} predicts growth in \texttt{r\Slash infinitewarfare}, but no other ecological interactions. 

The timepiece and Call of Duty ecological communities illustrate how subreddits with overlapping users can have relatively strong or weak forms of ecological interdependence.  Although both clusters are characterized by high degrees of user overlap and low average ecological interaction, the timepiece cluster has a dense competition-mutualism network while the call of duty network is sparse.

\subsubsection{Forecasting accuracy}
\label{sec:res.forecasting}

 

As described in §\ref{sec:mes.forecasting}, we test H2 by comparing two forecasting metrics of whether we have improved the 24-week forecast performance for all  subreddits which were assigned to clusters: root-mean-square-error (RMSE) and continuous ranked probability score (CRPS).
We find that VAR models including ecological interactions have forecasting performance superior to the baseline model in terms of both. The RMSE under the baseline model (0.84) is greater than the RMSE of the VAR models (0.75) and the CRPS of the baseline model (72,853) is also  greater than the CRPS of the VAR models (72,669).


\section{Threats to Validity}
\label{sec:limitations}
Our work is subject to several important threats to validity. First, we study only one platform hosting online groups and our results may not generalize to other platforms or time periods.
The method we propose for identifying ecological interactions between online groups has limitations common to all time series analysis of observational data. 
While our community ecology approach assumes that ecological interactions drive dynamics in the size of groups over time and cause groups to grow or decline, drawing causal conclusions using our method would depend on several untestable assumptions. For example, groups we do not consider---including groups on other platforms---could play a role in an ecological community accounted for in our models.  Potential omitted variables might also include additional time lags of group size. For simplicity we chose to use VAR(1) with a single time lag, but we hope future work will model more complex dynamics with additional lags.
Our vector autoregression models assumes that the error terms are trend stationary. This is a common assumption in time series analysis and is difficult to evaluate empirically  \cite{ives_estimating_2003}. 

Additional threats to validity stem from our use of algorithmic clustering to identify ecological communities.
While we choose clusters based on high degrees of user overlap and validate our clustering, we might have obtained different results if we had clustered in a different way. Additionally, our efforts to obtain clusters with a high silhouette coefficient led us to remove subreddits from our analysis. Thus, our results are not representative of Reddit overall, but only of those subreddits that were included in our analysis.

\section{Discussion}
\label{sec:discussion}
In the final chapter of their book on \textit{Building Successful Online Communities}, \citet{kraut_building_2012} advise managers of online groups to select an effective niche and beware of competition. However, these recommendations are based on little direct evidence from studies of online groups and offer almost no concrete steps that designer or group should take based on either piece of advice. Although further research into ecological interactions is needed to derive design principles, we provide a novel framework for online group managers to think about ecological constraints on group size. 
While intuition suggests that online group managers might seek out mutualistic relationships and avoid competitive ones, it is not clear whether another group is a competitor or mutualist. Our method provides a way for group managers to know. 

We presented two studies with the purpose of introducing our community ecology framework and comparing it to past work using population ecology.
In Study A, we found support for H1 showing---as predicted by density dependence theory---that overlap density has an $\cap$-shaped association with subreddit growth.
Subreddits with moderate overlap density in our data declined less than subreddits with either very low or very high overlap density. According to population ecology theory, this suggests that high-density environments are competitive and as a result are less conducive to growth than medium-density environments.

Surprisingly, this seems to contrast with our results in Study B. We studied the diversity of ecological communities using vector autoregression models of group size over time to infer networks of ecological interactions. 
We found that ecological communities of subreddits are typically mutualistic and that these mutualistic interactions are stronger on average than competitive ones.
These findings corroborate recent qualitative studies arguing that multiple online communities about the same topic exist because they provide different and complementary benefits \cite{teblunthuis_no_2022}, such as those provided by combinations of small and large subreddits \cite{hwang_why_2021}.

We also found a diversity of ecological dynamics among clusters of subreddits with high degrees of user overlap. We found ecological communities that are mutualistic, competitive, that mix the two, or that have few significant ecological interactions at all. 
This explains the puzzling set of empirical results from prior work about the relationship between overlap density and outcomes like growth, decline and survival  \cite{wang_impact_2012, zhu_impact_2014, zhu_selecting_2014}.
These studies have measured the density of an online group's niche in terms of its overlap in participants or topics. 
Although we find support for the relationship between density and growth predicted by density dependence theory, our analysis of ecological communities suggests that degrees of overlap may have little to do with whether two groups are mutualists or competitors. 
Instead, we argue that density can be associated with growth when a platform provides a hospitable environment to build online communities sharing certain topics or membership bases. 
Yet, when conditions change and these topics lose popularity or membership bases migrate off a platform \cite{fiesler_moving_2020}, this density may lead to decline. 

For example, differing environmental conditions of Wikis and Usenet groups might explain why user overlap was associated with the survival of wikis \cite{zhu_impact_2014} but with the decline of Usenet groups \cite{wang_impact_2012}. Wikia was a young and growing platform during \citepos{zhu_impact_2014} data collection period when the growth of groups may have been limited by knowledge of how organize and build a wiki; perhaps this knowledge was provided by overlapping experienced members. 
Usenet was in decline during \citepos{wang_impact_2012} study period and this may have created competition over increasingly scarce members. 

Future work should seek to explain when two online groups will be mutualists or competitors. 
Long-held understandings of ecological interactions in evolutionary theory suggest that, as we find, mutualism will be more common than competition \cite{aldrich_organizations_2006}. 
Competition is unlikely to persist because it decreases survival, but mutualistic relationships are likely to endure because they increase the parties' survival \cite{kropotkin_mutual_2012}.
In this line of theory, groups might avoid competition by adopting specialized roles in their ecological communities, a dynamic known as resource partitioning  \cite{carroll_concentration_1985}. For example, the competition among real estate subreddits observed in Figure \ref{fig:comp.network} may happen because of insufficient specialization.  
By contrast, mental health support groups like those observed in Figure §\ref{fig:comp.network} appear to have specialized purposes or roles. 


Online groups may use multiple platforms with distinctive affordances for different purposes \cite{kiene_technological_2019}. 
Since our VAR method relies only on time series data to infer ecological interactions, it can be applied to study ecological communities spanning social media platforms. 
While we focus on relationships between groups sharing a platform, one can apply our concepts and methods to how understand higher levels of social organization emerge from interdependent systems of technologies and users on social media platforms. 

\section{Conclusion}

An ecological explanation for the success of online groups looks beyond internal mechanisms to understand how different groups influence each other's growth or decline.
Prior research has investigated competition and mutualism among online groups with overlapping users and topics using the population ecology framework \cite{wang_impact_2012, zhu_impact_2014, zhu_selecting_2014}, yet has not provided a way to infer competitive or mutualistic interactions among related groups.
We introduce the community ecology framework as a complementary perspective to population ecology. 
By inferring competition-mutualism networks directly from time-series data, our community ecology approach helps resolve the empirical tensions raised by prior work and reveals that most interactions within clusters of highly overlapping subreddits are mutualistic. 
Our methods provide a foundation for future work investigating related online groups.  

\section{Ethics, Impact and Competing Interests}

The intended broader impact of this work is to improve the design and management of how online communities by advancing understanding of overlapping communities.  It is conceivable that this work may contribute to potential harms, such as if it is used to organize socially harmful online communities.  We hope and believe any negative consequences of this work will be outweighed by positive and productive ones. We are considerate of the ethics of studies such as ours, a large-scale analysis of publicly available social media communication and behavior in that the individuals whose traces we analyze are unlikely to anticipate their data will be used in such a way. That said, because our analysis aggregates these activities to such a degree that no individual is exposed to scrutiny, so we believe that the resulting harms are minimal. We have no competing financial interests in this work.

\bibliography{ms}

\end{document}